\documentclass[journal,twoside,web]{ieeecolor}
\usepackage{lcsys}
\usepackage{cite}
\usepackage{amsmath,amssymb,amsfonts, mathtools}
\usepackage{algorithmic}
\usepackage{graphicx}
\usepackage{textcomp}
\def\BibTeX{{\rm B\kern-.05em{\sc i\kern-.025em b}\kern-.08em
    T\kern-.1667em\lower.7ex\hbox{E}\kern-.125emX}}
\newcommand{\un}{$\;\cdot \; 10^6\; \text{m}^3$}

\begin{document}
\newtheorem{lemma}{Lemma}
\newtheorem{theorem}{Theorem}
\newtheorem{definition}{Definition}
\newtheorem{assumption}{Assumption}
\newtheorem{corollary}{Corollary}
\newtheorem{remark}{Remark}
\newtheorem{algorithm}{Algorithm}
\newtheorem{conjecture}{Conjecture}

\title{Toward Scalable Risk Analysis for Stochastic Systems Using Extreme Value Theory}
\author{Evan Arsenault, Yuheng Wang, and Margaret P. Chapman, \IEEEmembership{Member, IEEE} \vspace{-10mm}
\thanks{Manuscript received March 18, 2022; revised May 26, 2022;
accepted June 15, 2022. This research is supported by the Edward S. Rogers Sr. Department of Electrical and Computer Engineering (ECE), University of Toronto, and the Natural Sciences and Engineering Research Council of Canada (NSERC) Discovery Grants Program, [RGPIN-2022-04140]. Cette recherche a \'{e}t\'{e} financée par le Conseil de recherches en sciences naturelles et en g\'{e}nie du Canada (CRSNG). }
\thanks{The authors are with the Edward S. Rogers Sr. Department of ECE, University of Toronto, 10 King's College Road, Toronto, Ontario M5S 3G8 Canada.
Contact email: {\tt\small evan.arsenault@mail.utoronto.ca}.}
}

\maketitle
\thispagestyle{empty}

\begin{abstract}
We aim to analyze the behaviour of a finite-time stochastic system, whose model is not available, in the context of more rare and harmful outcomes. Standard estimators are not effective in making predictions about such outcomes due to their rarity.
Instead, we use \emph{Extreme Value Theory} (EVT), the theory of the long-term behaviour of normalized maxima of random variables. We quantify risk using the \emph{upper-semideviation} $\rho(Y) \coloneqq E(\max\{Y - \mu,0\})$ of an integrable random variable $Y$ with mean $\mu \coloneqq E(Y)$. $\rho(Y)$ is the risk-aware part of the common mean-upper-semideviation functional $\varphi_\lambda(Y) \coloneqq \mu + \lambda \rho(Y)$ with $\lambda \in [0,1]$. To assess more rare and harmful outcomes, we propose an EVT-based estimator for $\rho(Y)$ in a given fraction of the worst cases. We show that our estimator enjoys a closed-form representation in terms of the popular \emph{conditional value-at-risk} functional. In experiments, we illustrate the extrapolation power of our estimator using a small number of i.i.d. samples ($<$50). Our approach is useful for estimating the risk of finite-time systems when models are inaccessible and data collection is expensive.
The numerical complexity does not grow with the size of the state space.
\end{abstract}

\begin{IEEEkeywords}
Extreme Value Theory, Model-free risk estimation and analysis, Risk-aware stochastic systems\vspace{-3mm}
\end{IEEEkeywords}

\section{Introduction}
\label{sec:introduction}
\IEEEPARstart{T}{he} study of risk-aware systems has been growing in popularity due to its nuanced interpretations and assessments of uncertainties \cite{bauerlerieder, borkar2014risk, shen2014risk, van2015distributionally, milleryang, bauerle2020minimizing, huang2020stochastic, mpctacsubmission, kose2021risk}. A \emph{risk-aware} system is a dynamical system whose performance or safety is assessed in a manner that is sensitive to the possibility or severity of rare harmful outcomes. It is common to assess performance or safety using an expectation $E(Y)$, where $Y$ is a random cost, or a maximum deterministic cost $\sup\{c_i : i \in \mathcal{K}\}$, where $c_i$ is a real number indexed by $i \in \mathcal{K}$.

However, an expected cost is not designed to quantify more rare and harmful outcomes. Such outcomes are critical to assess when systems operate under uncertainties and have safety concerns. A maximum deterministic cost may be too conservative for some applications and is not suitable if $c_i$ is unbounded. Fortunately, one can define criteria that quantify many more characteristics of a random cost $Y$, such as: 
\begin{itemize}
    \item A weighted sum of the mean and variance,
    \item A quantile at a given probability level, and
    \item An expected exceedance above a threshold $\eta \in \mathbb{R}$, i.e., $E(\max\{Y - \eta,0\})$.
\end{itemize}

\noindent A popular criterion is the conditional value-at-risk (CVaR), which represents the expectation of $Y$ conditioned on $Y$ being larger than a particular quantile \cite[Th. 6.2]{shapiro2009lectures}. CVaR has been useful for defining performance or safety objectives \cite{milleryang, mpctacsubmission} and constraints \cite{borkar2014risk, van2015distributionally} for control systems. Another standard criterion is the mean-upper-semideviation (MUSD) $\varphi_\lambda(Y) \coloneqq \mu + \lambda \rho(Y)$ with $\lambda \in [0,1]$, which is a weighted sum of the mean $\mu \coloneqq E(Y)$ and the upper-semideviation $\rho(Y)$ defined by \vspace{-1mm}
\begin{equation}\label{myrho}
    \rho(Y) \coloneqq E(\max\{Y - \mu,0\}).
\end{equation}

\vspace{-2mm}The above criteria are examples of risk functionals. A \emph{risk functional} is a map from a space of random variables to the extended real line. CVaR and MUSD are real-valued \emph{coherent} risk functionals on the space of integrable random variables \cite[Ex. 6.16, Ex. 6.20]{shapiro2009lectures}. A real-valued coherent risk functional can be expressed as a robust expectation with respect to a particular family of distributions \cite[Th. 6.6]{shapiro2009lectures}, connecting risk analysis with distributionally robust optimization. Recent empirical evidence suggests that MUSD can provide protection against value function approximation errors \cite[Sec. 7]{kose2021risk}, reflecting its coherence. The expressive nature of risk functionals and the distributionally robust attributes of coherent ones suggest broad applicability to systems operating under uncertainties in practice. 

However, current approaches to analyze and optimize risk-aware non-linear systems suffer from scalability challenges. Most approaches are based on dynamic programming (DP) \cite{bauerlerieder, mpctacsubmission, bauerle2020minimizing} or Q-learning \cite{shen2014risk, huang2020stochastic}. These algorithms were developed originally to minimize an expected cumulative cost for a Markov decision process (MDP). When reformulating one of the above algorithms for a risk-aware setting, the new algorithm \emph{inherits the scalability issues} of the original algorithm. It is well-known that DP cannot apply to high-dimensional state spaces without function approximations. Q-learning typically involves extensive exploration, finite state spaces, and infinite time horizons \cite{shen2014risk, huang2020stochastic}. These conditions need not apply when sampling is moderately expensive, the state space is continuous, or analysis on a finite time horizon is needed. A recent scalable approach for risk analysis combines temporal difference learning with value function approximation and quantifies risk using a composition of coherent risk functionals \cite{kose2021risk}. The approach does not require extensive exploration but involves a finite-state infinite-time MDP \cite{kose2021risk}. Similar to \cite{kose2021risk}, we focus on the analysis of risk rather than its optimization. However, we adopt statistical tools from Extreme Value Theory (EVT). 

EVT is the study of the long-term behaviour of normalized maxima of random variables. It has been useful for examining extreme events in hydrology \cite{towler2010modeling}, seismology \cite{shcherbakov2019forecasting}, and disease transmission \cite{lim2020time}. This theory offers tools to extrapolate beyond the available data to estimate the upper tail of a distribution \cite{de2006extreme, longin2016extreme}. Theoretical connections between EVT and hitting time statistics for discrete-time systems have been established \cite{freitas2010hitting}. EVT has been applied to compute properties of chaotic systems, for example, the dimensions of an invariant measure \cite{caby2019extreme}. Statistical applications of EVT include estimating extreme quantiles \cite{blanchet2020distributionally}, CVaR \cite{mcneil1999extreme, troop2019risk, deo2020optimizing}, and extreme probabilities \cite[Ch. 4.4]{de2006extreme}. A distributionally robust approach for extreme quantile estimation with sensitivity to modeling errors has been proposed \cite{blanchet2020distributionally}. An EVT-based formula for CVaR \cite{mcneil1999extreme} has been applied to a multi-armed bandit problem \cite{troop2019risk}. Deo and Murthy have estimated CVaR and an associated gradient by combining tools from EVT and importance sampling \cite{deo2020optimizing}. This letter lies in the intersection of EVT, risk analysis, and stochastic systems. \vspace{-3mm}

\subsection{Problem formulation}\label{problemform}
We consider a stochastic system, whose dynamical model is not available, operating on a discrete finite time horizon of length $N \in \mathbb{N}$, a natural number. The state space $\mathbb{S}$, observation space $\mathbb{Y}$, and control space $\mathbb{C}$ are
Euclidean spaces. The sample space $\Omega \coloneqq (\mathbb{S} \times \mathbb{Y} \times \mathbb{C})^{N} \times \mathbb{S} \times \mathbb{Y}$ contains all possible trajectories. An element $\omega \in \Omega$ takes the form $\omega = (x_0,y_0,u_0,\dots,x_{N-1},y_{N-1},u_{N-1},x_N,y_N)$, where $x_t$ is a state, $y_t$ is an observation, and $u_t$ is a control. The $\sigma$-algebra $\mathcal{F}$ of interest is the Borel $\sigma$-algebra on $\Omega$ \cite[Def. 7.6]{bertsekas2004stochastic}. An initial condition $x \in \mathbb{S}$, a control policy $\pi$, and the (unknown) dynamical model induce a probability measure $P^\pi_x$ on $(\Omega, \mathcal{F})$ \cite[Prop. 7.28]{bertsekas2004stochastic}; an example of $\pi$ is an output feedback controller, which maps prior observations to a current control. Let $Y$ be a random variable on the probability space $(\Omega, \mathcal{F},P^\pi_x)$, where a realization of $Y$ is a cost that may be incurred by the system's behaviour. We assume that $Y$ is integrable, but its distribution is not known. Our goal is to estimate the risk of $Y$, provided that i.i.d. samples are available but not plentiful. We will write $P$ in lieu of $P_x^\pi$ for brevity.

While risk can be quantified in different ways, we consider the upper-semideviation $\rho(Y)$ \eqref{myrho}, that is, the risk-aware term of the coherent MUSD functional. $\rho(Y)$ enjoys the intuitive interpretation of being the expected exceedance of $Y$ above its mean. While there is literature about EVT-based estimation of CVaR \cite{mcneil1999extreme, troop2019risk, deo2020optimizing}, we have not found literature about EVT-based estimation of $\rho(Y)$. Moreover, we focus on estimating the upper-semideviation \emph{in a fraction of the largest (i.e., most harmful) realizations of $Y$}. This estimation task is especially difficult because such realizations are observed rarely. Yet, estimation in the context of rare and harmful outcomes is critical for systems with safety concerns. Hence, it is our focus.\vspace{-3mm}

\subsection{Contribution}
We propose an EVT-based estimator for the upper-semideviation in a fraction of the worst cases. Our theoretical contribution is to prove that the estimator admits a closed-form representation in terms of CVaR (Theorem \ref{thm3}). The proof requires the construction of a random variable $Y_\theta$ with useful properties (Lemmas \ref{distprop}--\ref{geninverse}). We construct $Y_\theta$ so that its distribution resembles the distribution of $Y$ in the upper tail under assumptions from Extreme Value Theory. In experiments, we demonstrate the ability of our estimator to perform well when the quantity of data is limited (Sec. \ref{num}, Fig. \ref{exp1_results}). We also showcase our approach using data of total overflow volumes from combined sewer systems throughout Canada \cite{statcanadacsos}. This letter initiates a new avenue for scalable risk analysis in data-sparse applications. \vspace{-3mm}

\section{Preliminaries}\label{evtsect}
\emph{Notation}. $\mathbb{N} \coloneqq \{1,2,\dots\}$ is the set of natural numbers, $\mathbb{R}$ is the set of real numbers, and $\mathbb{R}_{+} \coloneqq (0,+\infty)$. 
$Y \in L^1(\Omega, \mathcal{F}, P)$ or $Y \in L^1$ means that $Y$ is an integrable random variable on $(\Omega, \mathcal{F}, P)$, i.e., $E(|Y|)$ is finite. The function $1_{A} : \Omega \rightarrow \{0,1\}$ is the indicator on $A \in \mathcal{F}$. 
If $F$ is a distribution function, $z^* \coloneqq \sup\{z \in \mathbb{R} : F(z) < 1\}$ is its \emph{right endpoint}. 
For an $\mathbb{R}$-valued function $f$, we define $\tilde{f} \coloneqq 1 - f$. If $S$ is a metric space, $\mathcal{B}_{S}$ is the Borel $\sigma$-algebra on $S$, and 
$\text{int}(S)$ is the interior of $S$. Abbreviations: i.i.d. = independent identically distributed; a.e. = almost everywhere or almost every; w.r.t. = with respect to; d.o.f. = degrees of freedom.

Since EVT is not well-known in control theory, it is necessary to summarize some fundamentals, which we adopt from \cite{de2006extreme} and \cite{longin2016extreme}. Let $(Z_i)_{i \in \mathbb{N}}$ be an i.i.d. sequence of random variables defined on a probability space $(\Omega,\mathcal{F},P)$ with distribution function $F$.
We do not know $F$ or $P$. The partial maximum for $m \in \mathbb{N}$ is defined by
   $ M_m \coloneqq \max\{Z_1,Z_2,\dots,Z_m\}$.
Since $M_m$ converges in probability to the right endpoint $z^*$ of $F$, 
a normalization of $M_m$ can be useful for revealing characteristics of $F$. 
\begin{definition}[$F \in \mathcal{D}(G_\gamma)$]\label{maxdom}
Suppose that there exist sequences $(a_m)_{m \in \mathbb{N}} \subseteq \mathbb{R}_{+}$ and $(b_m)_{m \in \mathbb{N}} \subseteq \mathbb{R}$ such that 
\begin{equation}\label{keyrelation}
    \lim_{m \rightarrow +\infty} P\left(\Big\{\omega \in \Omega : \textstyle\frac{M_m(\omega) - b_m}{a_m} \leq z \Big\}\right) = G_\gamma(z)
\end{equation}
for every continuity point $z$ of $G_\gamma$, where $G_\gamma$ is a non-degenerate distribution function. Then, we say that $F$ belongs to the \emph{maximum domain of attraction of $G_\gamma$}, i.e., $F \in \mathcal{D}(G_\gamma)$.
\end{definition}

$G_\gamma$ being non-degenerate means that it does not correspond to a point mass. 
There are many examples of $F$ that satisfy Definition \ref{maxdom}, including: Pareto, Burr, Fr\'{e}chet, $t$-Student, Cauchy, Log-gamma, Uniform on $(0,1)$, Beta, Exponential, Logistic, Gumbel, Normal, Lognormal, and Gamma \cite{longin2016extreme}. The \emph{extreme value index} $\gamma$ is a qualitative measure for tail ``heaviness,'' i.e., how fast the tail of $F$, provided that $F \in \mathcal{D}(G_{\gamma})$, decays to zero \cite[p. 63]{longin2016extreme}.

The next theorem provides an equivalent characterization for $F \in \mathcal{D}(G_{\gamma})$, offering an approximation for the upper tail of $F$. For convenience, we define the interval $\mathcal{J}_\gamma$ by $\mathcal{J}_\gamma \coloneqq [0,+\infty)$ if $\gamma \geq 0$; $\mathcal{J}_\gamma \coloneqq [0,-1/\gamma)$ if $\gamma < 0$.
Moreover, we define the function $\phi_\gamma : \mathcal{J}_\gamma \rightarrow (0,1]$ by
\begin{equation}\label{phigamma}
    \phi_\gamma(z) \coloneqq  \begin{cases}(1 + \gamma  z)^{-1/\gamma}, & \text{ if } \gamma \neq 0, \\ \exp(-z), & \text{ if } \gamma = 0. \end{cases}
\end{equation}
$\tilde{\phi}_\gamma \coloneqq 1 - \phi_\gamma$ corresponds to the \emph{Generalized Pareto} distribution \cite[Eq. (4.6)]{longin2016extreme}.
\begin{theorem}\label{dehanslimit}
\cite[Th. 1.1.6, Part 4]{de2006extreme}: $F \in \mathcal{D}(G_{\gamma})$ for some $\gamma \in \mathbb{R}$ if and only if there is an $\mathbb{R}_+$-valued function $g$ s.t. 
\begin{equation}\label{19}
    \lim_{s \uparrow z^*} \frac{1 - F( s+ x  g(s))}{1 - F(s)} = \phi_\gamma(x), \quad x \in \mathcal{J}_\gamma,
\end{equation}
where $s \uparrow z^*$ means that $s$ approaches $z^*$ from below.
\end{theorem}

For brevity, we use the notations $g_s \coloneqq g(s)$ and $\tilde{F} \coloneqq 1-F$. Motivated by \eqref{19}, the following heuristic is commonly used, e.g., see \cite[pp. 65--66]{de2006extreme}, to approximate the upper tail of $F$ above some sufficiently large threshold $s \in \mathbb{R}$:
\begin{equation}\label{heu}
\tilde{F}(z) \approx \tilde{F}(s) \cdot \phi_{\gamma}( (z-s)/g_s ), \quad z \geq s.
\end{equation}

The subsequent lemma formalizes the tail approximation \eqref{heu} and uses the following definitions. We define a parameter vector $\theta$ by
\begin{equation}\label{mytheta}
    \theta \coloneqq \{k, m, \gamma, s, g_s\},  
\end{equation}
where $k \in \mathbb{N}$ and $m \in \mathbb{N}$ with $k < m$, $\gamma \in \mathbb{R}$, $s \in \mathbb{R}$, and $g_s \in \mathbb{R}_+$. If $F \in \mathcal{D}(G_{\gamma})$, then $\gamma$ is an extreme value index, $s$ is a threshold, and $g_s = g(s)$, where $g$ comes from Theorem \ref{dehanslimit}. 
In addition, we define the interval $\mathcal{I}_\theta$ by \vspace{-1mm}
\begin{align}\label{Itheta}
\mathcal{I}_\theta & \coloneqq  \begin{cases} [s,+\infty), & \text{if }\gamma \geq 0, \\ [s,s - g_s/\gamma), & \text{if }\gamma < 0. \end{cases}
\end{align}
\begin{lemma}[Formalized tail approximation]\label{lemma1}
Let $F \in \mathcal{D}(G_\gamma)$ for some $\gamma \in \mathbb{R}$, and 
let $\epsilon \in \mathbb{R}_+$ and $z \in \mathcal{I}_\theta$ be given.
If $z^* \in \mathbb{R}$, then there exists a $\delta_{\epsilon,z} \in \mathbb{R}_+$ such that \vspace{-1mm}
\begin{equation}\label{13}
   |\tilde{F}(z) - \tilde{F}(s) \cdot \phi_{\gamma}((z-s)/g_s)| < \epsilon \tilde{F}(s)
\end{equation}
for every $s \in ( - \delta_{\epsilon,z} + z^*, z^*)$.
Otherwise, if $z^* = +\infty$, then $\exists \; r_{\epsilon,z} \in \mathbb{R}$ such that \eqref{13} holds for every $s \in (r_{\epsilon,z}, z^*)$.
\end{lemma}
\hspace{-2mm}\begin{proof}
First, note that $F(z) < 1$ for every $z \in (-\infty,z^*)$ as a consequence of $z^*$ being the right endpoint of the distribution function $F$. Then, the result follows from applying the definition of the limit \cite[Def. 4.33, p. 98]{rudin1964principles} to \eqref{19}.
\end{proof}

Lemma \ref{lemma1} formalizes the tail approximation \eqref{heu} by providing a pointwise bound \eqref{13} on the magnitude of the approximation error. Next, we will apply the tail approximation to estimate the upper-semideviation in a fraction $\alpha$ of the worst cases, to be denoted by $\rho_\alpha(Y)$. We will derive a closed-form expression for an EVT-based estimator for $\rho_\alpha(Y)$ using first principles from probability theory.\vspace{-3.5mm} 

\section{Estimating extremal upper-semideviation}\label{estsec}
Let $Y \in L^1(\Omega,\mathcal{F},P)$ be a random cost, whose distribution function, 
$F_Y(y) \coloneqq P(\{\omega \in \Omega : Y(\omega) \leq y\})$ with $y \in \mathbb{R}$,
is not known. As we have motivated in Sec. \ref{problemform}, our focus is estimating the upper-semideviation $\rho(Y)$ \eqref{myrho} when considering a given fraction $\alpha \in (0,1)$ of the largest realizations of $Y$. We call this quantity the \emph{extremal upper-semideviation} of $Y \in L^1$ at level $\alpha$, which we define by \vspace{-1mm}
\begin{equation}\label{rhoalpha}
    \rho_\alpha(Y) \coloneqq \int_{\{\omega \in \Omega : Y(\omega) \geq v_{\alpha}(Y) \}} \max\{Y - \mu,0 \} \; \mathrm{d}P,
\end{equation}
where $v_{\alpha}(Y) \in \mathbb{R}$ is a threshold (to be specified). The meaning of $\rho_\alpha(Y)$ \eqref{rhoalpha} is the expected exceedance of $Y$ above the mean in a fraction $\alpha$ of the worst cases. In the definition of $\rho_\alpha(Y)$ \eqref{rhoalpha}, the integral is over a subset of the sample space $\Omega$ corresponding to a fraction of the largest realizations of $Y$. In contrast, in the definition of $\rho(Y)$ \eqref{myrho}, the integral is over the entire sample space $\Omega$.
The threshold $v_{\alpha}(Y)$ is the \emph{value-at-risk} of $Y$ at level $\alpha$ \cite[Sec. 6.2.3]{shapiro2009lectures}, i.e., 
\begin{equation}\label{myvar}
    v_{\alpha}(Y) \coloneqq \inf\{z \in \mathbb{R} : F_Y(z) \geq 1 - \alpha \}, \quad \alpha \in (0,1).
\end{equation}

To estimate $\rho_\alpha(Y)$ \eqref{rhoalpha}, we will apply properties of CVaR.
The CVaR of $Y \in L^1$ at level $\alpha \in (0,1)$ is defined by
\begin{equation}\label{cvardefdef}
    c_{\alpha}(Y) \coloneqq  \underset{\tau \in \mathbb{R}}{\inf} \Big( \tau + \textstyle\frac{1}{\alpha} E(\max\{Y-\tau,0\}) \Big)
\end{equation} 
as per \cite[Eq. (6.22)]{shapiro2009lectures}. 
A useful fact is that a minimizer of the right side of \eqref{cvardefdef} is $v_{\alpha}(Y)$ \cite[p. 258]{shapiro2009lectures}, 
and therefore,
\begin{equation}\label{cvarmin}
     c_{\alpha}(Y) =   v_{\alpha}(Y) + \textstyle\frac{1}{\alpha} E(\max\{Y-v_{\alpha}(Y),0\}).
\end{equation}

We will use a data set $\{y_{i,m} : i = 1,2,\dots,m\} \subset \mathbb{R}$ of size $m \in \mathbb{N}$ that is sampled independently from $Y$ and satisfies 
\begin{equation}\label{data}
y_{1,m} \leq \dots \leq y_{m-k,m} \leq \dots \leq y_{m-1,m} \leq y_{m,m}
\end{equation}
with $k \in \mathbb{N}$ and $k < m$. The notation $\mu_m \coloneqq \frac{1}{m}\sum_{i=1}^m y_{i,m}$ denotes the sample mean of the data \eqref{data}. To describe the data formally, let $(Y_i)_{i \in \mathbb{N}}$ be i.i.d. random variables defined on $(\Omega,\mathcal{F},P)$ with distribution function $F_Y$. For a given $m \in \mathbb{N}$, the notation $y_{i,m}$ denotes a realization of $Y_{j}$ for a particular $j \in \{1,2,\dots,m\}$, i.e., $y_{i,m} = Y_{j}(\omega)$ for some $\omega \in \Omega$. A data set $\{y_{i,m} : i = 1,2,\dots,m\}$ corresponds to an $\omega \in \Omega$, where we choose each index $i$ to satisfy the inequalities in \eqref{data}. 
\vspace{-3mm}

\subsection{Typical empirical estimator for $\rho_\alpha(Y)$}
A typical estimator for $\rho_\alpha(Y)$ \eqref{rhoalpha} is computed using the $k+1$ largest samples for some large enough $k \in \mathbb{N}$:\vspace{-1mm}
\begin{equation}\label{standardest}
    \hat{\rho}_{\alpha,k,m} \coloneqq \frac{1}{m} \sum_{i=m-k}^m \max\{y_{i,m} - \mu_m,0\},
\end{equation}
 %
where 
$y_{m-k,m}$ is an approximation for $v_{\alpha}(Y)$ \eqref{myvar}. 
%
$\hat{\rho}_{\alpha,k,m}$ \eqref{standardest} is not designed to represent the upper tail of $F_Y$ when the number $m$ of samples is limited; we will illustrate limitations of $\hat{\rho}_{\alpha,k,m}$ numerically in Sec. \ref{num}. 
We aim to estimate $\rho_\alpha(Y)$ \eqref{rhoalpha} using a data set \eqref{data} in a manner that is sensitive to the burden of collecting numerous samples and the challenge of observing large realizations of $Y$ in practice. \vspace{-3mm}

\subsection{EVT-based estimator for $\rho_\alpha(Y)$}
Here, we propose an EVT-based estimator for $\rho_\alpha(Y)$ \eqref{rhoalpha} that enjoys a closed-form representation in terms of CVaR. Our tactic is to construct a random variable $Y_\theta$, whose extremal distribution approximates the extremal distribution of $Y$. 

We suppose that an experiment has been conducted, providing a data set \eqref{data}. We let $\theta$ \eqref{mytheta} be a parameter vector, which has been estimated using this data set. There are different ways to estimate $\theta$ \cite{de2006extreme}, and we will illustrate one way in Sec. \ref{num}. Motivated by the analysis of Lemma \ref{lemma1}, we define for $z \in \mathbb{R}$
\begin{equation}\label{Ftheta}
   \hspace{-.5mm} F_{\theta}(z) \coloneqq \begin{cases} 0, & \hspace{-1.5mm}\text{if } z < s, \\ 1 - \frac{k}{m}  \phi_{\gamma}\hspace{-.8mm}\left(\frac{z-s}{g_s}\right)\hspace{-.8mm}, & \hspace{-1.5mm}\text{if }z \in \mathcal{I}_\theta, \\ 1, & \hspace{-1.5mm}\text{if } z \geq s - g_s/\gamma, \; \gamma < 0,\end{cases}
\end{equation}
with $s \coloneqq y_{m-k,m}$. The factor $\frac{k}{m}$ is an estimate for $\tilde{F}_Y(s) = 1 - F_Y(s)$, where the empirical distribution 
$\hat{F}_m(x) \coloneqq \frac{1}{m} \sum_{i=1}^m 1_{\{Y_{i} \leq x\}}$ 
is used because $F_Y$ is not known.
$F_{\theta}$ \eqref{Ftheta} satisfies key properties, as described below. 
\begin{lemma}[$F_{\theta}$ is a distribution function]\label{distprop}
For any $\theta$ \eqref{mytheta}, $F_{\theta} : \mathbb{R} \rightarrow [0,1]$ \eqref{Ftheta} is non-decreasing and right-continuous with $\underset{z \rightarrow -\infty}{\lim} F_{\theta}(z) = 0$ and $\underset{z \rightarrow +\infty}{\lim} F_{\theta}(z) = 1$.
\end{lemma}

The proof relies on applying the right-continuous and non-increasing properties of $\phi_\gamma$ \eqref{phigamma}, which we omit in the interest of space.
Lemma \ref{distprop} guarantees that $F_{\theta}$ is the distribution function of some random variable \cite[p. 209]{ash1972probability}, which we denote by $Y_\theta$. Using the canonical construction, 
we define $Y_\theta$ on the probability space $(\mathbb{R},\mathcal{B}_{\mathbb{R}},P_\theta)$, where $P_\theta$ is the Lebesgue-Stieltjes measure corresponding to $F_{\theta}$ and $Y_\theta(z) \coloneqq z$ for every $z \in \mathbb{R}$ \cite[p. 209]{ash1972probability}.
By Lemma \ref{lemma1}, under the assumption that $F_Y \in \mathcal{D}(G_\gamma)$ for some $\gamma \in \mathbb{R}$,
the approximation $F_{\theta} \approx F_Y$ is valid on $\mathcal{I}_\theta$ \eqref{Itheta}. 
Then, $Y_\theta$ and $Y$ have similar distributions in the upper tails. 
Next, we provide a sufficient condition for $Y_\theta \in L^1$, which will be useful for evaluating the CVaR of $Y_\theta$. 

\begin{lemma}[$Y_\theta \in L^1$]\label{lemma3}
Let $Y_\theta$ be a random variable with distribution function $F_\theta$ \eqref{Ftheta}. 
If 
$\gamma < 1$, then $Y_\theta \in L^1(\mathbb{R},\mathcal{B}_{\mathbb{R}},P_\theta)$.
\end{lemma}
\hspace{-3mm}\begin{proof}
It suffices to show that 
    $\int_{\mathbb{R}} |Y_\theta| \; \mathrm{d}P_{\theta} < +\infty$.
Take $b \in \mathbb{R}$ such that $b < s$, and define $Y_\theta' \coloneqq Y_\theta - b$. Therefore,
\begin{equation}
\label{lemma3eq1}
    P_\theta(\{z \in \mathbb{R}: Y_\theta'(z) > x\}) = 1 - F_\theta(x + b), \;\;\;x \in \mathbb{R}.
\end{equation}
Since $F_\theta(b) = 0$ by \eqref{Ftheta}, $Y_\theta'$ is positive a.e. w.r.t. $P_\theta$. By
a generalized tail integral formula \cite[Prop. 6.24]{folland1999real}, we have that
\begin{equation}\label{41}
   \int_{\mathbb{R}} |Y_\theta'| \; \mathrm{d}P_{\theta} = \int_0^{+\infty} 1 - F_\theta(x + b) \; \mathrm{d}x.
\end{equation}
Since $\gamma < 1$, the integral in the right side of \eqref{41} is finite by lengthy but standard calculus calculations. Then, we apply Minkowski's Inequality \cite[p. 183]{folland1999real} to $Y_\theta = Y_\theta' + b$ to conclude that
   $ \int_{\mathbb{R}} |Y_\theta| \; \mathrm{d}P_{\theta} \leq \int_{\mathbb{R}} |Y_\theta'| \; \mathrm{d}P_{\theta} + |b| < +\infty$.
\end{proof}

Given a data set \eqref{data} with sample mean $\mu_m \in \mathbb{R}$ and a parameter vector $\theta$ \eqref{mytheta}, we propose an estimator for $\rho_{\alpha}(Y)$ \eqref{rhoalpha} as follows:
\begin{equation}\label{myestimator}
    \hat{\rho}_{\alpha,\theta} \coloneqq \int_{A_{\alpha,\theta}} \max\{z - \mu_m, 0\} \; \mathrm{d}P_{\theta}(z), \quad \alpha \in (0, \textstyle\frac{k}{m}),
\end{equation}
where $A_{\alpha,\theta}$ is an interval defined by
\begin{equation}\label{Atheta}
    A_{\alpha,\theta} \coloneqq [v_{\alpha}(Y_\theta), +\infty), \quad \alpha \in (0, \textstyle\frac{k}{m}),
\end{equation}
and $v_{\alpha}(Y_\theta) \in \mathbb{R}$ is the value-at-risk of $Y_\theta$ at level $\alpha \in (0,1)$, 
\begin{equation}\label{myvarforest}
    v_{\alpha}(Y_\theta) \coloneqq \inf\{z \in \mathbb{R} : F_{\theta}(z) \geq 1 - \alpha \}. 
\end{equation}

\begin{remark}[About the definition of $\hat{\rho}_{\alpha,\theta}$]
The quantity $\hat{\rho}_{\alpha,\theta}$ \eqref{myestimator} concerns the realizations in the upper tail of $F_\theta$ \eqref{Ftheta}, and the quantity $\rho_\alpha(Y)$ \eqref{rhoalpha} concerns the realizations in the upper tail of $F_Y$. We define $\hat{\rho}_{\alpha,\theta}$ by \eqref{myestimator} because the upper tail of $F_\theta$ is a theoretically justified approximation for the upper tail of $F_Y$ under the assumptions of Lemma \ref{lemma1}. 
The next lemma specifies properties of $F_{\theta}(v_{\alpha}(Y_\theta))$ and $v_{\alpha}(Y_\theta)$.
\end{remark}

\begin{lemma}\label{geninverse}
Let $\alpha \in (0,\frac{k}{m})$, where $k \in \mathbb{N}$ and $m \in \mathbb{N}$ are from a given $\theta$ \eqref{mytheta} with $k < m$. Then, the following statements hold: (a) $F_{\theta}(v_{\alpha}(Y_\theta)) = 1 - \alpha$, and (b) $v_{\alpha}(Y_\theta) \in \text{int}(\mathcal{I}_{\theta})$.
\end{lemma}
\hspace{-3mm}\begin{proof}
To show (a), note that $F_{\theta}(v_{\alpha}(Y_\theta)) \geq 1 - \alpha$ with equality if and only if $1-\alpha$ is in the range of $F_{\theta}$ \cite[Lemma 21.1 (ii)]{van2000asymptotic}. Also, note that $F_{\theta}|_{\mathcal{I}_{\theta}} = [1 - \frac{k}{m},1)$ by direct evaluation of \eqref{Ftheta}. Since $0<\alpha < \frac{k}{m}<1$, we have $1>1 - \alpha > 1 - \frac{k}{m}$, and thus, $1 - \alpha$ is in the range of $F_{\theta}$. 

To show (b), we must show that $v_{\alpha}(Y_\theta) > s$, and in the case of $\gamma < 0$, we must also show that $v_{\alpha}(Y_\theta) < s - \frac{g_s}{\gamma}$. First, assume that $v_{\alpha}(Y_\theta) \leq s$. Since $F_\theta$ is non-decreasing (Lemma \ref{distprop}), $F_\theta(v_{\alpha}(Y_\theta)) \leq F_\theta(s)$. Moreover, we have that $F_{\theta}(v_{\alpha}(Y_\theta)) = 1 - \alpha > 1 - \frac{k}{m} = F_\theta(s)$ by applying Part (a), the assumption $\alpha < \frac{k}{m}$, and the definition of $F_\theta$ \eqref{Ftheta}, respectively. This is a contradiction, completing the proof for $\gamma \geq 0$. The case of $\gamma < 0$ follows from similar arguments.
\end{proof}

Next, we show that $\hat{\rho}_{\alpha,\theta}$ \eqref{myestimator} enjoys a closed-form representation in terms of the CVaR of $Y_\theta$ at level $\alpha$, which we denote by $c_{\alpha}(Y_\theta)$. 
Recall that $v_{\alpha}(Y_\theta)$ \eqref{myvarforest} is the value-at-risk of $Y_\theta$ at level $\alpha$, and $\mu_m$ is the sample mean of the data \eqref{data}. 

\begin{theorem}[Closed-form expression for $\hat{\rho}_{\alpha,\theta}$]\label{thm3}
Assume the conditions of Lemma \ref{lemma3}, and let the data $\{y_{i,m} : i = 1,2,\dots, m\}$ \eqref{data} be given. Suppose that $\theta$ \eqref{mytheta} has been estimated from the data with $s \coloneqq y_{m-k,m}$, and
%
%
let $0 <\alpha < \frac{k}{m} < 1$. If $v_{\alpha}(Y_\theta) \geq \mu_m$, then 
   $ \hat{\rho}_{\alpha,\theta} = \alpha \left(c_{\alpha}(Y_\theta) - \mu_m \right)$.
\end{theorem}
\hspace{-3mm}\begin{proof}
Lemma \ref{lemma3} guarantees that $Y_\theta \in L^1(\mathbb{R},\mathcal{B}_{\mathbb{R}},P_\theta)$. Thus, the CVaR of $Y_\theta$ at level $\alpha \in (0,1)$ satisfies
\begin{equation}\label{cvar_var}
     c_{\alpha}(Y_\theta) =   v_{\alpha}(Y_\theta) + \frac{1}{\alpha} \int_{\mathbb{R}} \max\{z - v_{\alpha}(Y_\theta), 0 \} \; \mathrm{d}P_{\theta}(z)
\end{equation}
by utilizing the argument underlying \eqref{cvarmin}.
For brevity, we use the notation $(x)^+ \coloneqq \max\{x,0\}$ for every $x \in \mathbb{R}$. By applying the definition of $A_{\alpha,\theta}$ \eqref{Atheta} and the assumption $v_{\alpha}(Y_\theta) \geq \mu_m$, it follows that $1_{A_{\alpha, \theta}}(z) \cdot (z - \mu_m)^{+} = $
\begin{equation}\begin{aligned}
   1_{A_{\alpha, \theta}}(z) \cdot (z - v_{\alpha}(Y_\theta))^{+}  + 1_{A_{\alpha, \theta}}(z) \cdot (v_{\alpha}(Y_\theta) - \mu_m)
\end{aligned}\end{equation}
for every $z \in \mathbb{R}$. By re-expressing $\hat{\rho}_{\alpha,\theta}$ \eqref{myestimator} and noting that a sum of non-negative Borel-measurable functions can be integrated term by term 
\cite[Cor. 1.6.4]{ash1972probability}, it holds that
\begin{align*}
    \hat{\rho}_{\alpha,\theta}  =  \int_{\mathbb{R}} 1_{A_{\alpha, \theta}}(z) \cdot (z - \mu_m)^{+} \; \mathrm{d}P_{\theta}(z)
     = \psi_{1,\theta} + \psi_{2,\theta},
     \end{align*}
     where $\psi_{1,\theta}$ and $\psi_{2,\theta}$ are defined by
     \begin{align}
    \psi_{1,\theta} & \coloneqq  \int_{\mathbb{R}} 1_{A_{\alpha, \theta}}(z) \cdot  (z - v_{\alpha}(Y_\theta))^{+} \; \mathrm{d}P_{\theta}(z), \label{psi1}\\
    \psi_{2,\theta} & \coloneqq  (v_{\alpha}(Y_\theta) - \mu_m) \cdot P_{\theta}(A_{\alpha, \theta}), \label{psi2}
\end{align}
%
respectively.
Since $A_{\alpha, \theta} = [v_{\alpha}(Y_\theta), +\infty)$ \eqref{Atheta}, we have that
\begin{equation}
    1_{A_{\alpha, \theta}}(z) \cdot  (z - v_{\alpha}(Y_\theta))^{+} = (z - v_{\alpha}(Y_\theta))^{+}, \quad z \in \mathbb{R},
\end{equation}
and therefore,
   $ \psi_{1,\theta} = \int_{\mathbb{R}}(z - v_{\alpha}(Y_\theta))^{+} \; \mathrm{d}P_{\theta}(z)$.
Next, we will simplify $\psi_{2,\theta}$ \eqref{psi2}. 
Using $A_{\alpha, \theta} = [v_{\alpha}(Y_\theta), +\infty)$, $F_\theta$ being a distribution function (Lemma \ref{distprop}), and $P_\theta$ being the corresponding Lebesgue-Stieltjes measure, we have that
\begin{align}\label{54}
    P_{\theta}(A_{\alpha, \theta}) = 1 - \underset{z \uparrow v_{\alpha}(Y_\theta)}{\lim} F_{\theta}(z)
\end{align}
by \cite[1.4.5 (9), p. 25]{ash1972probability}. Since $F_{\theta}$ \eqref{Ftheta} is continuous on $\text{int}(\mathcal{I}_{\theta})$, $v_{\alpha}(Y_\theta) \in \text{int}(\mathcal{I}_{\theta})$, and $F_\theta(v_{\alpha}(Y_\theta)) = 1-\alpha$ (Lemma \ref{geninverse}), we find that
\begin{equation}\label{55}
    \underset{z \uparrow v_{\alpha}(Y_\theta)}{\lim} F_{\theta}(z) = F_\theta(v_{\alpha}(Y_\theta)) = 1-\alpha.
\end{equation}
%
%
%
Using \eqref{54}--\eqref{55}, we simplify $\psi_{2,\theta}$ \eqref{psi2} as follows:
\begin{equation}\label{psi2.1}
    \psi_{2,\theta} =  (v_{\alpha}(Y_\theta) - \mu_m)  (1 - (1 - \alpha)) = \alpha  (v_{\alpha}(Y_\theta) - \mu_m).
\end{equation}
Using our simplifications for $\psi_{1,\theta}$ and $\psi_{2,\theta}$,
we conclude that
\begin{align}
    \hat{\rho}_{\alpha,\theta} 
    &= \alpha  \left(\frac{1}{\alpha}\int_{\mathbb{R}}(z - v_{\alpha}(Y_\theta))^{+} \; \mathrm{d}P_{\theta}(z) + v_{\alpha}(Y_\theta) - \mu_m \right) \nonumber\\
    &= \alpha(c_{\alpha}(Y_\theta) - \mu_m),
\end{align}
where we use \eqref{cvar_var} in the final line.
\end{proof}

\begin{remark}[Theorem \ref{thm3} assumptions]
Typically, $\alpha$ is small, e.g., less than 0.05, to emphasize rare high-consequence outcomes.  
Thus, we anticipate $v_{\alpha}(Y_\theta) \geq \mu_m$ to hold 
in applications. We will present an example in Sec. \ref{sysexnum}. 
\end{remark}

Subsequently, we use techniques from \cite{mcneil1999extreme, shapiro2009lectures} to provide expressions for $v_{\alpha}(Y_\theta)$ \eqref{myvarforest} and $c_{\alpha}(Y_\theta)$ \eqref{cvar_var}.
\begin{remark}[Expressions for $v_{\alpha}(Y_\theta)$ and $c_{\alpha}(Y_\theta)$]\label{mycor}
Assume the conditions of Theorem \ref{thm3}. 
Since $0 < \alpha < \frac{k}{m} < 1$, it holds that $v_{\alpha}(Y_\theta) \in \text{int}(\mathcal{I}_{\theta})$ by Lemma \ref{geninverse}. Since $F_\theta$ \eqref{Ftheta} is continuous and strictly increasing on $\text{int}(\mathcal{I}_{\theta})$, we invert $F_\theta$ on $\text{int}(\mathcal{I}_{\theta})$ to derive the following expression for $v_{\alpha}(Y_\theta)$:
\begin{equation}\label{var2}
    v_{\alpha}(Y_\theta)= \begin{cases}s + \frac{g_s}{\gamma}\left( \left(\frac{m \cdot \alpha}{k} \right)^{-\gamma} - 1\right), & \text{if }\gamma \neq 0, \\s - g_s \cdot \log\left(\frac{m \cdot \alpha}{k} \right), & \text{if } \gamma = 0. \end{cases}
\end{equation}
As well as $\alpha \in (0,1)$, we have that $Y_\theta \in L^1(\mathbb{R},\mathcal{B}_{\mathbb{R}},P_\theta)$ by Lemma \ref{lemma3}. 
Further, CVaR can be expressed as an average of the value-at-risk, $c_{\alpha}(Y_\theta) = \frac{1}{\alpha} \int_{1-\alpha}^1 v_{1-\tau}(Y_\theta) \; \mathrm{d}\tau$ \cite[Th. 6.2]{shapiro2009lectures}.
We use this expression with $v_{1-\tau}(Y_\theta)$ from \eqref{var2} and the assumption $\gamma < 1$ from Theorem \ref{thm3} to derive
\begin{equation}\label{cvar2}
    c_{\alpha}(Y_\theta) = (v_{\alpha}(Y_\theta) + g_s - \gamma \cdot s)  (1-\gamma)^{-1}. \vspace{2.5mm}
\end{equation}
\end{remark}

Theorem \ref{thm3} and Remark \ref{mycor} together provide a closed-form expression for $\hat{\rho}_{\alpha, \theta}$ \eqref{myestimator}, which we use for computation.  
\vspace{-2mm}

\section{Numerical experiments}\label{num}
We conduct two experiments regarding the estimation of $\rho_\alpha(Y)$ \eqref{rhoalpha}. First, we compare the performance of a typical estimator $\hat{\rho}_{\alpha,k,m}$ \eqref{standardest}
to our EVT-based estimator $\hat{\rho}_{\alpha,\theta}$ \eqref{myestimator} using six benchmark distributions. 
The second experiment uses data of combined sewer overflows in Canada \cite{statcanadacsos}. Our code is available from https://github.com/eArsenault/evt-control.

Given a data set \eqref{data}, we estimate a parameter vector $\theta = \{k, m, \gamma, s, g_s\}$ \eqref{mytheta} using the probability-weighted moment estimator (PWME). The PWME is consistent under appropriate conditions 
\cite[Th. 3.6.1]{de2006extreme},
simple to implement, and produces a value for $\gamma$ that is strictly less than 1. 
The last characteristic is useful in light of Lemma \ref{lemma3}. For a given data set, we use the following procedure with $\alpha = 0.01$ and $\frac{k}{m} \approx 0.10$:
\begin{enumerate}
    \item We choose $s$ to be an estimate for $v_{0.10}(Y)$ \eqref{myvar}. We use $s \coloneqq y_{\lceil 0.90 \cdot m \rceil,m}$, where $\lceil \cdot \rceil$ is the ceiling function \cite{troop2019risk}.
    \item We assign $k$ to be the cardinality of $\{y_{i,m}: y_{i,m} > s, i = 1,2,\dots,m\}$ so that $y_{m-k,m} = s$ \cite[p. 66]{de2006extreme}. 
    \item We set $\gamma$ and $g_s$ as per the PWME expressions, which are closed-form and provided by \cite[Eqs. (3.6.9), (3.6.10)]{de2006extreme}.
\end{enumerate}
\begin{remark}[Consistency discussion]
The PWME is consistent when there is a family $(Z_i)_{i \in \mathbb{N}}$ of i.i.d. random variables with distribution function $F \in \mathcal{D}(G_\gamma)$ \eqref{keyrelation} with $\gamma < 1$ such that the $m$th order statistics $Z_{1,m} \leq \dots \leq Z_{m-k,m} \leq \dots \leq Z_{m,m}$ satisfy $k \rightarrow +\infty$ and $\frac{k}{m} \rightarrow 0$ as $m \rightarrow +\infty$ \cite[Th. 3.6.1]{de2006extreme}. Our numerical procedure uses $m < 100$ samples because we are concerned with data-sparse applications. Further experiments could be designed for consistency in particular by considering significantly larger data sets, whose order statistics satisfy the limiting conditions in principle (e.g., choose $k$ to grow logarithmically with respect to $m$). Each benchmark distribution $F_Y$ satisfies $F_Y \in \mathcal{D}(G_\gamma)$ with $\gamma < 1$ (see below). 
\end{remark}
\subsubsection{Benchmark distributions}
We compare the performance of the estimators $\hat{\rho}_{0.01,k,m}$ \eqref{standardest} and $\hat{\rho}_{0.01,\theta}$ \eqref{myestimator} for $m \in \{20,21,22,\dots,99\}$ samples. For each $m$, we have conducted 10,000 runs of the following: 1) draw $m$ i.i.d. samples from a random variable $Y$; 2) estimate $\theta$ as described previously; and 3) compute the errors $\hat{\rho}_{0.01,k,m} - \rho_{0.01}(Y)$ and $\hat{\rho}_{0.01, \theta} - \rho_{0.01}(Y)$. We approximate the ground-truth value of $\rho_{0.01}(Y)$ using a Monte Carlo simulation with over 4 million samples.

\begin{figure*}[h!]
\centering
\includegraphics[scale=0.66]{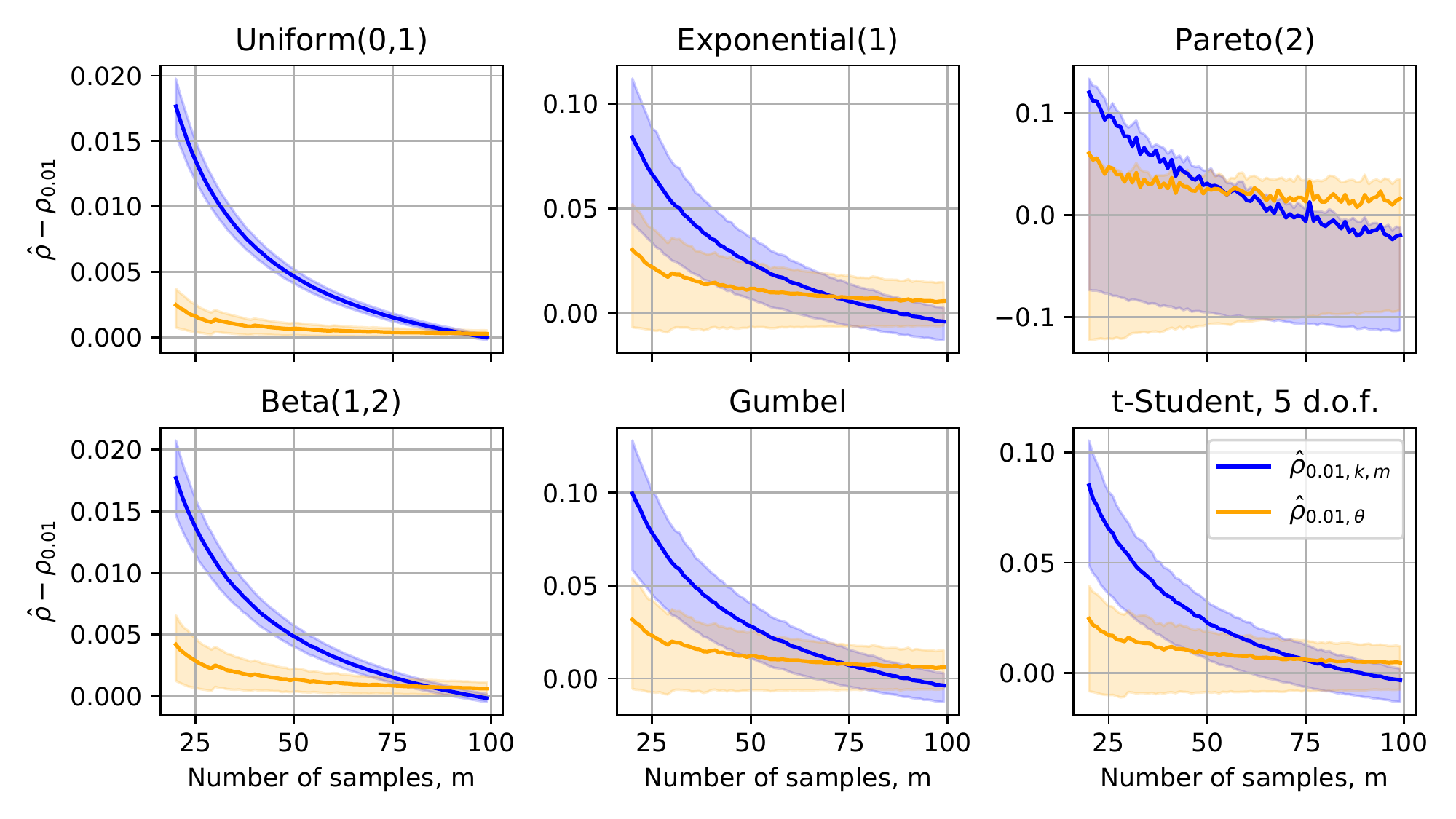}\vspace{-3mm}
\caption{The plots depict the errors, $\hat{\rho}_{0.01,k,m} - \rho_{0.01}(Y)$ (blue) and $\hat{\rho}_{0.01, \theta} - \rho_{0.01}(Y)$ (orange), versus the number of samples $m$. $\hat{\rho}_{0.01,k,m}$ \eqref{standardest} is a typical estimator, and $\hat{\rho}_{0.01,\theta}$ \eqref{myestimator} is our EVT-based estimator. The average error across 10,000 trials is plotted using the solid lines, and the coloured bands represent 50\% confidence intervals. The EVT-based estimator outperforms the typical estimator for smaller values of $m$.}\vspace{-5mm}
\label{exp1_results}
\end{figure*}

We consider six distributions for $Y$. 
The Pareto(2) and $t$-Student (5 d.o.f.) distributions have $\gamma = 0.5$ and $\gamma = 0.2$, respectively; the Exponential(1) and Gumbel distributions have $\gamma = 0$; the Uniform(0,1) and Beta(1,2) distributions have $\gamma < 0$ \cite{longin2016extreme}. 
Fig. \ref{exp1_results} presents the results. The EVT-based estimator $\hat{\rho}_{0.01, \theta}$ has lower average error versus the typical estimator $\hat{\rho}_{0.01,k,m}$ for smaller values of $m$. The error decreases as $m$ increases, and for sufficiently large $m$, the typical estimator is superior. 
These results support using the EVT-based estimator for risk analysis when a small number of samples is available (which is our focus). We illustrate one application next.

\subsubsection{System example}\label{sysexnum}
Monthly combined sewer overflow volumes in Canada during the years of 2013--2017 are available from \cite{statcanadacsos}. Each data point is the total volume of water (a mixture of stormwater and untreated wastewater) that was released into the environment during a particular month. (Combined sewers are present in older cities throughout North America.) The combined sewer network that underlies this data set is a vast hard-to-model dynamical system with safety requirements to protect environmental and public health. 
For the sake of illustration, we assume that the data during the spring and early summer months (March--June) is i.i.d., providing $m = 20$. We acknowledge that even data from a single month over consecutive years need not be i.i.d. due to climate trends.
We compute $\hat{\rho}_{0.01,k,m} = 0.59\text{\un}$ and $\hat{\rho}_{0.01, \theta} = 0.23\text{\un}$ with $k = 2$. We verify some assumptions from Theorem \ref{thm3}: $v_{0.01}(Y_{\theta}) = 26\text{\un} > \mu_m = 15\text{\un}$ and $\alpha = 0.01 < \frac{2}{20} = \frac{k}{m}$. We compute $\gamma = 0.87$, suggesting a heavy-tailed distribution. Interestingly, $\hat{\rho}_{0.01,k,m} > \hat{\rho}_{0.01, \theta}$, which also occurs for smaller values of $m$ in Fig. \ref{exp1_results}. Hence, the typical estimator may over-approximate the risk compared to the EVT-based estimator. \vspace{-3mm}

\section{Conclusions}\label{conc}\vspace{-1mm}
We have shown the ability of a new EVT-based estimator for a risk functional to perform well when data is limited. Our estimator was developed from measure-theoretic first principles, initiating a pathway for broader EVT-based risk analysis. We are in the process of deriving a consistency proof by applying theory from extreme quantile estimation \cite[Th. 4.3.1]{de2006extreme}; a key aspect is to provide interpretable conditions under which the EVT-based CVaR estimator \eqref{cvar2} is consistent using the facts that convergence in probability is preserved under continuous mappings and sums \cite{folland1999real,van2000asymptotic}. The derivation of a convergence rate for $\hat{\rho}_{\alpha,\theta}$ \eqref{myestimator} to the true value $\rho_{\alpha}(Y)$ \eqref{rhoalpha} is another vital direction. We anticipate rates to be distribution-dependent. We hypothesize faster rates when $F_Y$ is a Generalized Pareto distribution, based on \eqref{19} and an analogous result about convergence rates for \eqref{keyrelation} \cite{ROOTZEN1984219}. A study about convergence rates for the tail approximation \eqref{heu} for twice-differentiable invertible distribution functions \cite{raoult:hal-00693464} may also be useful. Ultimately, we hope to develop controllers that are sensitive to more rare and harmful outcomes for high-dimensional systems when data is limited. \vspace{-3mm}

\section*{Acknowledgment}\vspace{-1mm}
M.P.C. gratefully thanks Mr. Changrui Liu for discussions. \vspace{-7mm}

\bibliographystyle{IEEEtran}
\bibliography{references_new}

\end{document}